# An open question: Are topological arguments helpful in setting initial conditions for transport problems in condensed matter physics?

A. W. Beckwith

Department of Physics and Texas Center for Superconductivity and Advanced Materials at the University of Houston
Houston, Texas 77204-5005, USA

## ABSTRACT

The tunneling Hamiltonian is a proven method to treat particle tunneling between different states represented as wavefunctions in many-body physics. Our problem is how to apply a wave functional formulation of tunneling Hamiltonians to a driven sine-Gordon system. We apply a generalization of the tunneling Hamiltonian to charge density wave (CDW) transport problems in which we consider tunneling between states that are wavefunctionals of a scalar quantum field . We present derived *I-E* curves that match Zenier curves used to fit data experimentally with wavefunctionals congruent with the false vacuum hypothesis. The open question is whether the coefficients picked in both the wavefunctionals and the magnitude of the coefficients of the driven sine Gordon physical system should be picked by topological charge arguments that in principle appear to assign values that have a tie in with the false vacuum hypothesis first presented by Sidney Coleman.

Correspondence: A. W. Beckwith:    [projectbeckwith2@yahoo.com](projectbeckwith2@yahoo.com)

# INTRODUCTION

In this paper, I discuss whether topological charge arguments aid tunneling Hamiltonian (TH) formalism for charge density wave (CDW) transport. There are two alternatives conclusions. One is that the potential is set by the massive Schwinger model;[1] namely a driven Sine Gordon potential has coefficients that may or may not be set in absolute magnitude values by use of the Bogomol'nyi inequality as specified by Zee.[2] This is congruent with the false vacuum hypothesis.[3] The other conclusion is that the same magnitude of potential terms is an artifact of the experimental set up as was observed in 1985.[4] This indicated that the dominant $(1-\cos\phi)$ classical term is about 100 times larger than the coefficient in front of the driving quantum mechanical addition that has a scalar value proportional to $\phi^2$ in it.

In Section III, I argue that in using the tunneling Hamiltonian as proportional to current what is done with soliton- anti soliton (S-S') pairs is akin to coherent transfer of individual bosons. Thus, I used individual S-S' pairs in a thin wall approximation[3] with the distance specified to be inversely proportional to the coefficient $a$ in front of the Gaussian wave functional used in the tunneling Hamiltonian.

Experimental conditions could be the only decisive reason for the selection of the relative size of terms in our driven Sine Gordon potential — and the selection of the relative coefficient $a$ loosely necessitated by the S-S' pair being a Bosonic construction.[3] Else, the topological charge argument I am presenting may be useful as an extension of the physics presented by Mark Trodden[5] and Trodden et al[6] in viewing topological defects for condensed matter applications. I also assume that we can take the results of Frank Wilczek's[7] research into fractional quantum numbers for chiral fermions and assume that the charges of a S-S' pair cancel each other out on domain walls, permitting us to use the indicated topological construction. Finally, to conclude our research question in this paper, a driven Sine Gordon potential[8] has much of its potential contribution eliminated either due to purely experimental considerations or due to the vanishing of topological charge Q.[9] All this is provided assuming a S-S' nucleated pair due to the necessity of achieving a current-electric field (*I-E*) curve which fits experimental data sets as obtained in 1985 by Miller et al[4] in his NbSe3 experiments (quasi one dimensional metallic system at low temperatures with an applied electric field applied to the sample).

## II. SETTING A WAVE FUNCTIONAL IN TERMS OF GAUSSIAN PROBABILITY ALONE

When we work with classical formulation of matter states used in transport problems, we can make an analogy to the astrophysical literature. This permits us to look at how we may formulate states of matter nucleating from a purported vacuum state. Here, semi classical instanton approximations allows us to state that dominant contributions to the path integral come from metrics and Maxwell fields which are near the solutions that extremalize the Euclidian action and satisfy boundary conditions. So we may construct a wave function that that denotes the creation of a black hole via [10]

$$\psi_{inst} \equiv B \cdot \exp(-I_{inst}) \qquad (1)$$

where $B$ is a one loop contribution from quadratic fluctuations in the fields, $\partial^2 I$, and $I_{inst}$ is the classical action of the gravitational instanton that mediates the pair creation of black holes. I claim that we will look at Eq. 1 in terms of the stated action as an integration of a weakly coupled scalar field in a space of physical applications in which we wish to analyze our problem. For cosmology, the action can be in terms of a four- (or higher) dimensional space time. Whereas, in condensed matter, I will still be considering whether there exists a convenient geometry to simplify what is a least-action problem in setting up the condensed matter analogue to Eq. 1.

Is this the optimal configuration for condensed matter transport issues? Here, I consider what we can do when we work with a Gaussian wavefunctional as picked as an initial starting point for a given family of Hamiltonians. Assuming this is so, consider a Hamiltonian system for a sine Gordon style potential in several dimensions of the form[11]

$$H_O = \int_x \left[ \frac{1}{2} \cdot \Pi_x^2 + \frac{1}{2} \cdot (\partial_x \phi_x)^2 + \frac{1}{2} \cdot \mu^2 \cdot (\phi_x - \varphi)^2 - \frac{1}{2} \cdot I_0(\mu) \right] \quad (2)$$

One may obtain a ground-state wave functional of the form[11]

$$|0>^o = N \cdot \exp\left\{ -\int_{x,y} (\phi_x - \varphi) \cdot f_{xy} \cdot (\phi_y - \varphi) \right\} \quad (3)$$

where I have, due to higher order terms[11] in a perturbing potential, $H_1$

$$\frac{\partial^2 \cdot V_E}{\partial \cdot \phi^a \cdot \partial \cdot \phi^b} \propto f_{xy} \quad (4)$$

as this becomes equivalent to a coupling-term between the different branches of this physical system. Wu et al[11] used a multi-dimensional ground state anzatz. I restricted the analysis to quasi-one-dimensional cases, in which one would be able to observe a ground-state that looks like:

$$\Psi = |0>^0 \equiv c \cdot \exp(-\alpha \cdot \int dx [\phi - \phi_C]^2) \quad (5)$$

where I defined, as was also done Wu et al,[11]

$$\phi_C \equiv {}^o<0|\phi_x|0>^o \tag{5a}$$

where the $|0>$ is for a ground ( or a vacuum ) state and Eq. 5 is for a standard model physics trajectory with field evaluated at $x \equiv (x^1,....,x^D)$ being a position in $D$ dimensional space then being set to $D = 1$. I argue here that Eq. 5a choice of wave functional is due to variational considerations while fixing terms within the potential due to experimental considerations alone. However, I will present, in section III, an argument for a vanishing topological charge determination of the coefficients of the wavefunctional itself that also determines the relative size of the potential terms relative to each other. The alpha coefficient in Eq. 5a may be set by ad hoc experimental inputs, as well as by a topological argument we explore in the next section.

## III. TUNNELING HAMILTONIAN PROCEDURE USED IN OUR CDW EXAMPLE WITH COEFFICIENTS SET BY THE BOGOMOLNYI INEQUALITY

Traditional current treatments followed the Fermi golden rule for current density

$$J \propto W_{LR} = \frac{2 \cdot \pi}{\hbar} \cdot |T_{LR}|^2 \cdot \rho_R(E_R) \tag{6}$$

with $T_{LR}$ is very close to the form used by Tekeman.[12]

$$T_{mn} \equiv -\frac{\hbar^2}{2\mu} \int [\psi_0^* \nabla \psi_{mn} - \psi_{mn} \nabla \psi_0^*] \cdot dS \tag{7}$$

This is when identifying the $\psi_0$ as the initial wavefunction at the left hand side of a barrier, and $\psi_{mn}$ as the final wavefunction at the right hand side of a barrier. Note that Tekman[12] has extended the TH method to encompass more complicated geometries, such as the tips used in scanning tunneling microscopy and that his formulation is usually applied for a potential barrier with two turning points. What I contribute additionally is to notice that when the matrix elements $T_{kq}$ are small, the current through the barrier is calculated using linear response theory; this is found to be proportional to $|T|^2$ for quasiparticle tunneling, as suggested by Eq. 1. One should note that this may be used to describe coherent, Josephson-like tunneling of either Cooper pairs of electrons or boson-like particles such as superfluid [4] He atoms. In this case, the supercurrent goes linearly with the effective matrix element for transferring a pair of electrons or transferring a single boson, as shown rather elegantly in Feynman's derivation[13] of the Josephson current-phase relation. This means a current density proportional to $|T|$ rather than $|T|^2$ since tunneling, in this case, would involve coherent transfer of individual bosons (first-order) rather than pairs of fermions.[3] In this case of a current density proportional to $|T|$, I will be able to use the Bogomol'yi inequality[2,3] in order to isolate a Gaussian contribution to the wavefunctional states used in our field theoretic tunneling Hamiltonian. This allowed as a starting point a generalized wave functional

$$\psi \propto c \cdot \exp\left(-\beta \cdot \int L\, d\tau\right) \tag{8}$$

in a functional current we derived as being of the form

$$J \propto T_{if} \tag{9}$$

when

$$T_{if} \cong \frac{\hbar^2}{2\mu} \int \left( \Psi^*_{initial} \frac{\delta^2 \Psi_{final}}{\delta\phi(x)_2} - \Psi_{final} \frac{\delta^2 \Psi^*_{initial}}{\delta\phi(x)_2} \right) \vartheta(\phi(x) - \phi_0(x)) \wp\,\phi(x) \tag{5}$$

and where

$$c_2 \cdot \exp\left(-\alpha_2 \cdot \int d\tilde{x}\, [\phi_T]^2\right) \cong \Psi_{final} \tag{10}$$

and

$$c_1 \cdot \exp\left(-\alpha_1 \cdot \int dx [\phi_F]^2\right) \equiv \Psi_{initial} \tag{11}$$

where the $\alpha_2 \cong \alpha_1$ and $\phi_F \equiv <\phi>_1 \cong$ *very small value* as well as having in CDW $\phi_T \cong \phi_{2\pi} \equiv 2 \cdot \pi + \varepsilon^+$. These values for the phase showed up in the upper right hand side of Fig. 1a (as well in Fig. 1b) and represent the decay of the false vacuum hypothesis that I found to be in tandem with the Bogomil'nyi inequality.[2,3] As mentioned in a prior paper,[2] this allows presenting a change in energy levels to be inversely proportional to the distance between a S-S' pair

$$\alpha_2 \equiv \Delta E_{gap} \equiv \alpha \approx L^{-1} \tag{12}$$

I also found that in order to have a Gaussian potential in our wavefunctionals that we needed to have[2,3]

$$\frac{(\{\})}{2} \equiv \Delta E_{gap} \equiv V_E(\phi_F) - V_E(\phi_T) \tag{13}$$

for potentials of the form (generalization of the extended sine Gordon model potential)

$$V_E \cong C_1 \cdot (\phi - \phi_0)^2 - 4 \cdot C_2 \cdot \phi \cdot \phi_0 \cdot (\phi - \phi_0)^2 + C_2 \cdot (\phi^2 - \phi_0^2)^2 \tag{14}$$

I had a Lagrangian[2,3] modified to be (due to the Bogomil'nyi inequality)

$$L_E \geq |Q| + \frac{1}{2} \cdot (\phi_0 - \phi_C)^2 \cdot \{\} \tag{15}$$

with topological charge $|Q| \to 0$ and with the Gaussian coefficient found in such a manner as to leave one with wave functionals[2] generalized for charge density transport

$$\Psi_f[\phi(\mathbf{x})]\bigg|_{\phi \equiv \phi_{Cf}} = c_f \cdot \exp\left\{-\int d\mathbf{x}\, \alpha\left[\phi_{Cf}(\mathbf{x}) - \phi_0(\mathbf{x})\right]^2\right\}, \qquad (16)$$

and

$$\Psi_i[\phi(\mathbf{x})]\bigg|_{\phi \equiv \phi_{Ci}} = c_i \cdot \exp\left\{-\alpha \int d\mathbf{x}\left[\phi_{ci}(\mathbf{x}) - \phi_0\right]^2\right\}, \qquad (17)$$

I will perform a change of basis argument for Eqs. 16 and 17, pertinent to the thin wall approximation for CDW S-S' pairs traversing a pinning gap and do it in a way which permits analyzing equation 5 in momentum space. However, first I need to explain the physics used in the selection of the basis function used for this transport problem.

## IV. EVALUATING THE TUNNELING HAMILTONIAN ITSELF AS A WAY TO GET A 'CURRENT' CALCULATION IN CHARGE DENSITY WAVES

Regardless of how the wave functionals are picked,[14] to present how to match experimental data sets via the functional methods addressed here, I can use either the topological vanishing of charge argument or straight experimental inputs in Eqs. 12 and 13 as put into a wave functional representation of tunneling.

In arXIV,[3] I obtained $T_{IF}$ as having an absolute magnitude of

$$|T_{IF}| \approx \cdot \frac{2}{2 \cdot m^*}\left(n_1^2 - \frac{n_1^4}{2}\right) \cdot C_1 \cdot C_2 \cdot \left(\cosh\left(2\sqrt{\frac{x}{2L}} - \sqrt{\frac{L}{2x}}\right)\right) \cdot e^{-\alpha \cdot L \cdot \left[n_1^2 \cdot \frac{L}{2x}\right]} \qquad (18)$$

where I assumed using a scaling of $\hbar \equiv 1$ and $n_1 \cong 1 - \varepsilon^+$ becomes[3]

$$|T_{IF}| \approx \frac{C_1 \cdot C_2}{m^*} \cdot \left( \cosh\left( 2\sqrt{\frac{x}{2L}} - \sqrt{\frac{L}{2x}} \right) \right) \cdot e^{-\alpha \cdot L \left[ \frac{L}{2x} \right]} \quad (19)$$

This is due to a complex valued integration that would vanish if the imaginary contribution of $T_{IF}$ were ignored. Note that the undetermined coefficients in Eq. 19 were determined by first a thin wall approximation to the basis wavefunctionals given in Eqs. 16 and 17 as well as picking a momentum presentation for evaluating our tunneling matrix Hamiltonian, with the momentum version of a F.T. of the thin wall approximation[3] being set by

$$\phi\left(k_{n_{a1}, a2}\right) \equiv \phi(k) \equiv \phi(k_n) = \sqrt{\frac{2}{\pi}} \cdot \frac{\sin(k_n L/2)}{k_n} \quad (19a)$$

and I also assume a normalization of the form [3]

$$C_i = \frac{1}{\sqrt{\int_0^{\sqrt{\frac{L^2}{2 \cdot \pi}}} \exp\left(-2 \cdot \{\ \}_i \cdot \phi^2(k_N)\right) \cdot d\phi(k_N)}} \quad (19b)$$

where for the different wavefunctionals I evaluate for $i = 1,2$ via the error function[15]

$$\int_0^{\sqrt{\frac{L^2}{2 \cdot \pi}}} \Psi_i^2 \cdot d\phi(k_N) = 1 \quad (19c)$$

$$\int_0^b \exp(-a \cdot x^2) dx = \frac{1}{2} \cdot \sqrt{\frac{\pi}{a}} \cdot \mathrm{erf}\left(b \cdot \sqrt{a}\right) \quad (19d)$$

This is leading to a current that is the magnitude of a residue calculation[3] where we have

$$T_{IF} \approx \int \frac{f(k_N)}{g(k_N - i \cdot (value))} \cdot dk_N \tag{20}$$

where the numerator $f$ and denominator $g$ are analytic complex valued function. $T_{IF}$ would be zero if we were not counting imaginary root contributions to the functional integral for our tunneling Hamiltonian. Note that the S-S' pairs will form a current. This will occur when we have condensed electrons tunneling through a pinning gap at the Fermi surface. In order to accelerate the CDW with respect to an electric field, we have a de facto non-zero current composed of S-S' pairs when $E_{DC} \geq E_T$. Note that the Bloch bands are tilted by an applied electric field when we have $E_{DC} \geq E_T$ leading to a S-S' pair as shown in Fig. 2a[16]. The slope of the tilted band structure is given by $e^* \cdot E$ and the separation between the S-S' pair is given by:

$$L = \left(\frac{2 \cdot \Delta_s}{e^*}\right) \cdot \frac{1}{E} \tag{21}$$

So then we have $L \propto E^{-1}$ [3,16]. When we consider a Zener diagram of CDW electrons with tunneling only happening when $e^* \cdot E \cdot L > \varepsilon_G$ where $e^*$ is the effective charge of each condensed electron and $\varepsilon_G$ being a pinning gap energy, I find that Fig. 3 permits writing [3]

$$\frac{L}{x} \equiv \frac{L}{\overline{x}} \cong c_v \cdot \frac{E_T}{E} \tag{22}$$

Here, $c_v$ is a proportionality factor included to accommodate the physics of a given spatial (for a CDW chain) harmonic approximation of

$$\bar{x} = \bar{x}_0 \cdot \cos(\omega \cdot t) \Leftrightarrow m_{e^-} \cdot a = -m_{e^-} \cdot \omega^2 \cdot \bar{x} = e^- \cdot E \Leftrightarrow \bar{x} = \frac{e^- \cdot E}{m_{e^-} \omega^2} \quad (22a)$$

Realistically, an experimentalist will have to consider that $L \gg \bar{x}$, where $\bar{x}$ is an assumed reference point an observer picks to measure where a S-S' pair is an assumed one-dimensional chain of impurity sites. This sets writing the given magnitude of $|T_{IF}|$ as directly proportional to a current formed of S-S' pairs, which is further approximated to be[3]

$$I \propto \tilde{C}_1 \cdot \left[ \cosh\left[ \sqrt{\frac{2 \cdot E}{E_T \cdot c_V}} - \sqrt{\frac{E_T \cdot c_V}{E}} \right] \right] \cdot \exp\left( -\frac{E_T \cdot c_V}{E} \right) \quad (23)$$

where

$$\tilde{C}_1 \equiv \frac{C_1 \cdot C_2}{m^*} \quad (23a)$$

which is a great refinement upon the phenomenological Zenier current[4] expression

$$I \propto G_P \cdot (E - E_T) \cdot \exp\left( -\frac{E_T}{E} \right) \text{ if } E > E_T \quad (24)$$

otherwise

**[Figure 2a, 2b about here]**

## V. CONCLUSION

Herein, I restrict myself to analyzing ultra-fast transitions of CDW, which is realistic and in sync with how wavefunctionals used are formed in part by the fate of the false vacuum hypothesis. The question we need to address is, whether it is sufficient to put strictly experimental inputs into Eqs. 16 and 17 in order to get the physics of Fig. 1 or whether the Gaussian wave functional put in as a trial ansatz is sufficient for given experimental conditions. Needless to say, when the Bogomol'nyi inequality approach is used, it would be prudent to find suitable domain wall constructions along the lines of Su et al[17] that would incorporate a periodic boundary construction, which would permit vanishing of topological charge as was done by Trodden,[5,6] Su,[17] and others.[7]

In a subsequent publication, I will explore remarkable similarities between what we have presented here and Lin's[18] expansion of Schwinger's work in electron-positron pair production, which we believe is physically significant. Regardless of the approach used to form the wavefunctionals, as seen in Eqs. 16 and 17, the pinning wall interpretation of tunneling for CDW permits construction of *I-E* curves that match experimental data sets in a manner that beforehand were merely Zenier curve fitting polynomial constructions; this exhibits important new physics that we believe are useful for an experimentally based understanding of transport problems in condensed matter physics. Should the S-S' picture be enhanced by the topological issues presented here, I should still attempt to keep the basic structure of instanton physics intact and continue to adhere to the least action principle according to standards Zee[2] and Javir Casahoran,[19] while keeping in mind what M. Bowick, A. De Filice, and M. Trodden[6] presented in their generalized arguments.

# FIGURE CAPTIONS

**Fig. 1a:** Evolution from an initial state $\phi_i[\ ]$ to a final state $\phi_f[\ ]$ for a double-well potential (inset) in a 1-D model, showing a kink-antikink pair bounding the nucleated bubble of true vacuum. The shading illustrates quantum fluctuations about the classically optimum configurations of the field $\phi_i = 0$ and $\phi_f(x)$, while $\phi_0(x)$ represents an intermediate field configuration inside the tunnel barrier

**Fig. 1b:** Fate of the false vacuum representation of what happens in CDW. This shows how we have a difference in energy between false and true vacuum values and how this ties in with our Bogomil'nyi inequality.

**Fig. 2a:** The above figures represents the formation of soliton-anti soliton (S-S') pairs along a chain. The evolution of phase is spatially given by

$$\phi(x) = [\tanh b(x-x_a) + \tanh b(x_b - x)]$$

**Fig. 2b:** Experimental and theoretical predictions of current values. The dots represent a Zenier curve fitting polynomial, whereas the blue circles are for the S-S' transport expression derived with a field theoretic version of a tunneling Hamiltonian

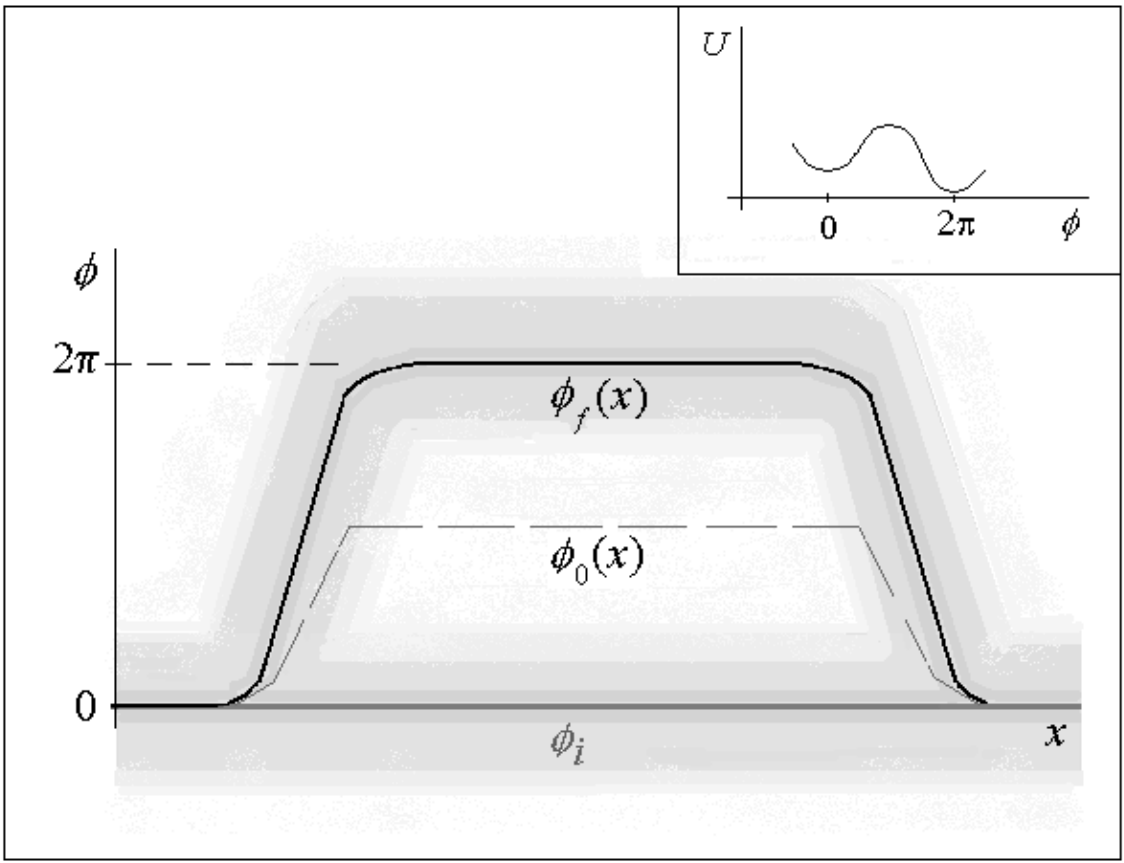

**Figure 1a**

**Beckwith**

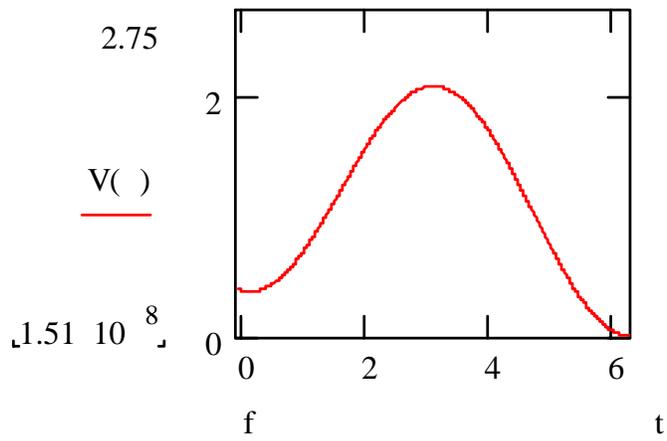

**Figure 1b**

**Beckwith**

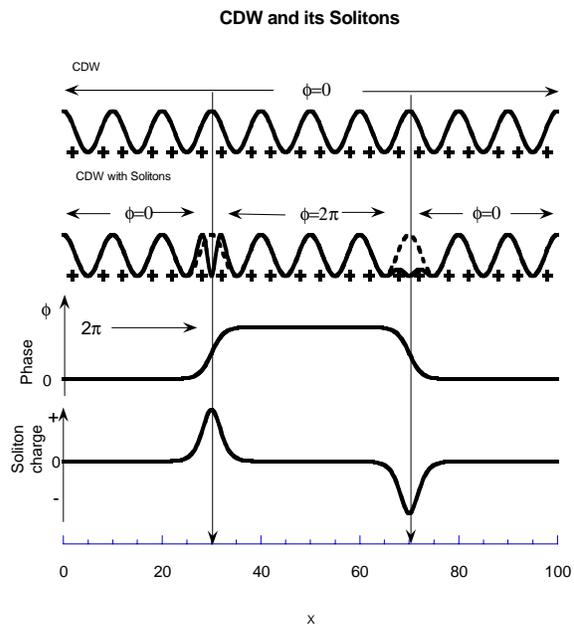

**Figure 2a**

**Beckwith**

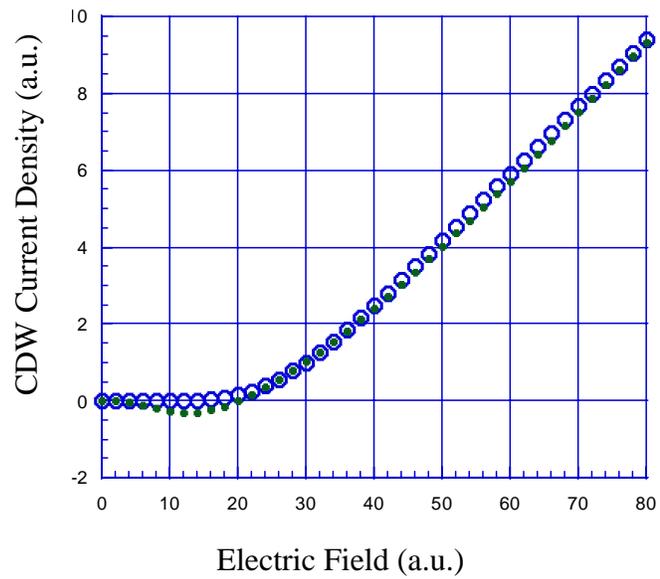

**Figure 2b**

**Beckwith**